\documentclass[conference]{IEEEtran}
\IEEEoverridecommandlockouts
\usepackage{cite}
\usepackage{amsmath,amssymb,amsfonts}
\usepackage{algorithmic}
\usepackage{graphicx}
\usepackage{textcomp}
\usepackage{xcolor}
\def\BibTeX{{\rm B\kern-.05em{\sc i\kern-.025em b}\kern-.08em
    T\kern-.1667em\lower.7ex\hbox{E}\kern-.125emX}}
\begin{document}

\title{DS2-Based Cross-Data-Space Interoperability for Precision Agriculture \thanks{This study was conducted in the context of the DS2 project (DataSpace, DataShare 2.0). The project has received funding from the EU Horizon Europe Research and Innovation Programme under grant agreement No. 101135967.}
}

\author{
\IEEEauthorblockN{
Katerina Kyriakou\IEEEauthorrefmark{1}\IEEEauthorrefmark{2},
Ilias Syrigos\IEEEauthorrefmark{1}\IEEEauthorrefmark{2},
Ioannis Moutsinas\IEEEauthorrefmark{3},
Panagiotis Tzimotoudis\IEEEauthorrefmark{1}\IEEEauthorrefmark{2},
Thanasis Korakis\IEEEauthorrefmark{1}\IEEEauthorrefmark{2}
}

\IEEEauthorblockA{
\IEEEauthorrefmark{1}
\textit{Department of Electrical and Computer Engineering,}\\
\textit{University of Thessaly, Greece}
}

\IEEEauthorblockA{
\IEEEauthorrefmark{2}
\textit{Centre for Research and Technology Hellas (CERTH), Greece}
}

\IEEEauthorblockA{
\IEEEauthorrefmark{3}
\textit{School of Agricultural Sciences,}\\
\textit{University of Thessaly, Greece}
}

\IEEEauthorblockA{
\{aikyriakou@uth.gr, ilsirigo@uth.gr, iomoutsi@uth.gr, tzimotou@uth.gr, korakis@uth.gr\}
}
}

\maketitle

\begin{abstract}
Despite the strategies of modern precision agriculture to leverage the integration of legacy agricultural systems, the challenges of IoT data fragmentation, farmers’ sovereignty preservation, and limited interoperability still persist. This paper presents our work, conducted within the Horizon Europe DS2 project (DataSpace, DataShare 2.0), that applies an interoperability-oriented framework supporting participants of different agricultural data spaces to share data products and services under secure, sovereign, and transparent methods. The suggested methodology follows a layered reference architecture, where each layer consists of independent operational modules that facilitate the inter-sector data exchange between DigiAgro and AgroScience Data Spaces. The result of this work is an automated ecosystem for sharing diverse farm IoT measurements, satellite images and metrics, weather forecasts, and analytics services across distinct data spaces, aiming to generate accurate recommendations on crop practices, such as irrigation schedules, that farmers and agronomists will rely on to increase crop production while maintaining sustainability. 
\end{abstract}


\section{Introduction}

Precision Agriculture applications are increasingly dependent on data coming from multiple and independently managed sources. Internet of Things (IoT) devices deployed in fields are constantly generating measurements related to soil and microclimate conditions, while weather services, crop-imaging systems, and agricultural analytics platforms provide complementary data that drive the implementation of farm management decision-support systems. Specifically, the combination of these data can support applications such as irrigation planning, crop-growth monitoring, yield estimation, disease detection, demand forecasting, and extreme-weather events assessment. However, combining these resources is currently being constrained due to the fragmentation of data across farms, research organisations, technology providers, and industry, and the reluctance of stakeholders who are concerned about the potential misuse of often commercially sensitive information, such as field conditions, production practices, and expected yields \cite{su151813746}. Data providers, therefore, demand complete control over which organisations can discover, request, and access their data and how the data may subsequently be used. On the other hand, data consumers require detailed metadata of the available datasets as well as explicit and well-defined contractual terms in order to determine whether they are suitable for specific analytical purposes. As a result, agricultural data sharing extends beyond a simple technical integration problem to a challenge of trust and data sovereignty.

To overcome such limitations, the notion of Data Spaces has gained considerable attention by the European Union as an infrastructure that enables automated data sharing while ensuring privacy-preservation, sovereignty, and interoperability. Existing initiatives such as IDSA \cite{ids_reference_model}, Gaia-X \cite{gaiax_concept}, and Data Space Support Centre (DSSC) through its Data Space Blueprint have focused on establishing universal, secure, and accountable standards and protocols required to create European Data Spaces. Although the increasing use of data spaces has contributed to the optimization of existing agricultural systems through EU-funded projects, such as the Common European Agricultural Data Space \cite{CEADS}, valuable agricultural data may still remain isolated, limiting the potential for the establishment of novel business models in the smart farming sector.

The Data Space Interoperability Architecture (DSIA) addresses this limitation by introducing interoperability services through which participants of distinct data spaces can execute contract negotiations and subsequent data exchanges among them \cite{SOININEN2026112678}. DSIA follows an intermediary approach, in which a set of shared services and modules constitute the bridging layer between independent data spaces. Thus, there is no requirement for direct integration of technologies, as this intermediary layer handles the necessary operations of identity management, publication and discovery of data offerings, contract negotiation, policy enforcement and data exchange on behalf of the participating entities. Although DSIA defines the generic functions for cross-data-space interoperability, its instantiation depends on the specific use case. Precision Agriculture presents a relevant case, as decision-support workflows can greatly benefit by the integration of heterogeneous IoT data, weather, satellite imagery, and analytical data, while preserving the sovereignty and control of the farmers acting as data providers.

This paper presents an instantiation of DSIA, conducted within the Horizon Europe DS2 project (DataSpace, DataShare 2.0) \cite{DS2Project}, for data exchanges between two complementary agricultural data spaces operating in Northern and Central Greece: DigiAgro Data Space \cite{SYRIGOS2026102247} and AgroScience Data Space. The former provides field-level observations collected from IoT sensor networks, while the latter provides analytical capabilities for agronomists and data analysts along with third-party data streams related to weather forecasts and satellite images, and metrics. The workflow demonstrated in this paper transfers field-related environmental data from DigiAgro DS to an AgroScience DS analytical service that generates field-specific irrigation decision-support.

The contributions of this work are:

\begin{itemize}
    \item A set of functional, technical, and operational requirements for interoperable data sharing between the two agricultural data spaces.
    \item A mapping of these requirements to the concepts and capabilities of DSIA and its corresponding components.
    \item A feasibility assessment of the instantiated workflow based on the implementation of an irrigation decision-support service.
\end{itemize}

The remainder of this paper is organised as follows. Section \ref{sec:related} presents related literature on cross-data-space interoperability. Section \ref{sec:requirements} presents the requirements of the use case.  Section \ref{sec:DSIA} describes the DSIA-based architecture, while Section \ref{sec:AgroUseCase} offers the details of the deployment and implementation of the agricultural data ecosystem. Section \ref{sec:workflow} demonstrates a cross-data-space scenario in the form of an irrigation decision-support system and Section \ref{sec:conclusions} concludes this paper.

\section{Related Work} \label{sec:related}
A review of the existing literature indicates that while several concepts have been proposed for cross-data-space interoperability, end-to-end implementations have not been widely reported. Existing interoperability approaches are generally classified into collaborative, federated, and intermediary-based Data Spaces. Collaborative Data Spaces enable cross-domain data sharing through homogeneous interlinking layers for establishing connections across independent platforms, which introduces significant technical integration overhead. Federated Data Spaces establish a centralized governance authority responsible for enabling data exchange across heterogeneous environments, thus creating administrative bottlenecks. 

Existing initiatives have contributed to addressing specific aspects of cross-data-space interoperability, while a comprehensive architecture for data exchange remains an area of ongoing research. Several EU Horizon projects have explored these aspects including PLIADES \cite{PLIADES}, and CyclOps \cite{CYCLOPS}, addressing use cases such as AI-driven metadata brokers, and the automation of FAIR data pipelines. In our precision agriculture use case, we implement DSIA that addresses this gap by following an intermediary-based approach. This architecture relies on three-party collaboration agreements to enable sovereign cross-data-space discovery, usage policy enforcement, and automated data transactions with minimal deployment overhead through containerization.

\section{Use Case Requirements} \label{sec:requirements}
To support the design and implementation of an interoperable data ecosystem, it is essential to understand the involved stakeholders' needs and translate those into concrete system requirements that will provide the basis for the proposed framework. Thus, the requirements elicitation process in the context of the DS2 project was based on the collaboration with representatives from DigiAgro and AgroScience data spaces during several technical workshops that involved stakeholders such as software engineers, agronomists, data analysts, and data space operators.

The derived requirements were divided into three separate categories including functional, technical and operational requirements. Functional requirements define the compulsory capabilities and actions of the interconnected data ecosystem. Technical requirements determine the infrastructure, protocols and data constraints to offer interoperability in the cross-data-space system. Operational requirements impose the necessities in terms of data privacy, trust and governance. 

\subsection{Functional Requirements}
The functional requirements are defined as follows. \textbf{FR1 (Data product publication and discovery)} enables DigiAgro members to publish descriptions of agricultural data products and authorized AgroScience participants to discover these products. \textbf{FR2 (Cross-data-space data exchange)} supports the transfer of environmental observations between participants belonging to the two data spaces without requiring either data space to replace its existing infrastructure. \textbf{FR3 (Data validation and transformation)} detects malformed or incomplete sensor records and transforms valid observations into the schema required by the AgroScience analytical services. Finally, \textbf{FR4 (Decision-support result delivery)} makes the output of the decision-support services available to an authorized user of either data space or to an application in a machine-readable or visual form.
% \begin{itemize}
%     \item \textbf{FR1 - Data product publication and discovery.} The system must allow DigiAgro to publish descriptions of agricultural data products and must allow authorized AgroScience participants to discover these products.
%     \item \textbf{FR2 - Cross-data-space data exchange.} The system must support the transfer of environmental observations between participants belonging to the two data spaces without requiring either data space to replace its existing infrastructure.
%     \item \textbf{FR3 - Data validation and transformation.} The system should detect malformed or incomplete sensor records and transform valid observations into the schema required by the AgroScience analytical services.
%     \item \textbf{FR4 - Decision-support result delivery.} The system should make the output of the decision-support services available to an authorized user of either data spaces or to an application in a machine-readable or visual form.

% \end{itemize}

\subsection{Technical Requirements}
The technical requirements are defined as follows. Interoperability and security across data spaces are governed by six technical requirements. Standardized alignment with the Data Space Protocol is enforced by \textbf{TR1 (Interoperable data transaction protocol)} for negotiation, execution, logging, and data transfers. \textbf{TR2 (Identity and membership validation)} verifies certificates and active memberships across both data spaces. Access and usage conditions are evaluated dynamically via \textbf{TR3 (Machine-readable contracts and policies)}, while \textbf{TR4 (Data model interoperability)} supports mapping diverse schemas, ontologies, and semantic models. In addition, \textbf{TR5 (Secure service communication)} ensures authenticated and encrypted communications, and \textbf{TR6 (Modular deployment)} requires interoperability services to be packaged as containerized modules deployable alongside host infrastructure.

% \begin{itemize}
%     \item \textbf{TR1 - Interoperable data transaction protocol.} Standardized alignment with the Data Space Protocol must be ensured for contract negotiation, contract execution, logging and data transfers.
%     \item \textbf{TR2 - Identity and membership validation.} Identity validation must be supported across the two data spaces, by verifying certificates and validating active memberships of participants. 
%     \item \textbf{TR3 - Machine-readable contracts and policies.} The system must represent access and usage terms in a machine-readable format and must evaluate these conditions during data transactions.
%     \item \textbf{TR4 - Data model interoperability.} The system should support the mapping of diverse formats, data schemas, ontologies, and semantic interpretations between the two data spaces.
%     \item \textbf{TR5 - Secure service communication.} The system should ensure communication between participant services is authenticated and secure.
%     \item \textbf{TR6 - Modular deployment.} The required interoperability services that allow cross-data-space transactions should be containerized and deployable alongside the existing infrastructure.

% \end{itemize}

\subsection{Operational Requirements}
The operational requirements are defined as follows. \textbf{OR1 (Data sovereignty)} guarantees the protection of data owners' legal rights and access boundaries. Under \textbf{OR2 (Cross-data-space collaboration agreement)}, cross-boundary transactions can proceed only after an active tripartite agreement is established between data space operators and the DSIA operator. Finally, \textbf{OR3 (Legal and policy compliance support)} enforces regulatory compliance across participating organizations to align with their respective legal obligations.
% \begin{itemize}
%     \item \textbf{OR1 - Data sovereignty.} Data sovereignty must be ensured to provide protection of data owners' legal rights.
%     \item \textbf{OR2 - Cross-data-space collaboration agreement.} Cross-data-space transactions must take place only after a valid collaboration agreement exists between the data spaces (DigiAgro, AgroScience) operators and DSIA operator.
%     \item \textbf{OR3 - Legal and policy compliance support.} Regulatory compliance must be enforced for cross-sector data exchanges to ensure compliance with the participating organisations' legal and governance obligations.

% \end{itemize}

\section{Reference Architecture} \label{sec:DSIA}
%% Federation purpose (general)
DSIA focuses on providing interoperability, trustworthiness, and sovereignty for the facilitation of cross-data-space data exchange. The reference architecture leverages the capabilities of a trusted Operator acting as an intermediary component, which is responsible for enhancing security in inter-sector data sharing, while retaining the autonomy of existing data spaces. The Operator establishes collaboration agreements for rules among participating data spaces' authorities, and thus allows the participants of different data spaces to bridge the data sharing gap across independent ecosystems. Regarding trust verification, the Operator relies on existing data spaces as the main source of truth and extends it by performing security risk assessment. 

%% Federator-Operator role in agricultural
The reference architecture is tailored to tackle the multi-domain sharing limitations in agricultural systems posed by the fragmentation of field and environmental measurements, and the concerns of farmers for compromising privacy. The system follows a decentralized governance framework in which the Operator utilizes three-party collaboration agreements and provides federation services for the facilitation of cross-organisational agricultural data sharing. Integrating the intermediary operator addresses the complexities of data exchange transactions, while safeguarding sovereignty, automation, and minimized onboarding overhead. The framework consists of three autonomous sub-systems including the Interoperability Governance system, the Data Space Interoperability system, and the Service Deployment system, enforcing a separation of concerns rather than providing a monolithic governance authority.

To illustrate a workflow of the DSIA subsystems in precision agriculture, an irrigation decision-support scenario is examined. In this scenario, farmers participating in DigiAgro Data Space act as data providers offering IoT sensor telemetry, such as soil moisture and climate metrics, while agronomists and data analysts from a company from AgroScience Data Space utilize these field data to generate automated irrigation scheduling recommendations. Initially, the DigiAgro and AgroScience Data Space operators sign a three-party collaboration agreement with the DSIA Operator to establish trust through the Interoperability Governance system. Once participant memberships are verified, the Data Space Interoperability system handles the publishing and discovery of the farmer's offerings, enabling the agronomists to initiate contract negotiation by the DS2 Connectors. During data transfer, farmer-defined usage policies (e.g., time-based restrictions) are strictly enforced, while on-the-fly data pipelines transform raw measurements into a standardized format required by the irrigation model. Finally, the Service Deployment system manages the seamless deployment of the interoperability modules as containerized packages on the hosts legacy systems. 

%% DSIA reference architecture
% DSIA is designed to provide a system of modular components that is dependent on the utilization of the Operator for handling federation services related to Data Space actions. 
% The purpose is to provide an ecosystem where participants of different Data Spaces can perform automated contract agreements for data sharing without being hindered by the technical overhead of heterogeneous legacy infrastructures. 

\subsection{Interoperability Governance System}
The Interoperability Governance System activates technical processes related to operational, collaboration, membership, and conflict management aiming to legally bind different data spaces. It creates an administrative framework that validates collaboration agreements, issues member identities, handles participant registries, and defines operational rules. 

\subsection{Data Space Interoperability System}
The data space interoperability system establishes three core subsystems that focus on enabling secure data transfers, orchestrating data pipelines, and providing AI-driven support services to efficiently facilitate data sharing across different data spaces.

\subsubsection{Data Sharing}
The Data Sharing subsystem orchestrates the data transfer execution by extending the Eclipse implementation \cite{ECLIPSE} to provide an interoperable Connector, enhanced with supporting mechanisms for trustworthiness, data sovereignty, and traceability. The core DSIA Connector component is responsible for contract negotiation and enforcement, as well as for the bits transfer across data spaces. The complementary components manage identity provisioning and validation of data space members and DSIA services, publication and discovery of data products in a Catalog, and immutable logging and monitoring.

\subsubsection{Data Pipelines}
The Data Pipeline subsystem addresses the complexity that arises from the data heterogeneity of different data spaces in terms of ontologies, languages, and formats. It enables on-the-fly data transformations as a part of the data sharing transactions. The pipeline subsystem consists of the orchestrator for workflow planning and runtime controlling, the modules for interoperability related services in data sharing, and the module repository for storing and publishing the pipeline applications, such as transformation, curation, and quality assessment modules. 

\subsubsection{Supporting Services}
The Supporting Services subsystem prioritizes providing services designed to guide users in the integration of their existing data spaces in the interoperability ecosystem by navigating through the potential complexities of the legacy systems. These services include AI-driven tools that enable 
automated data offering preparation and creation, AI chatbots that perform effective querying and filtering on data offering discovery, and risk analysis tools that verify trustworthiness and legal compliance of data transactions. Finally, sovereignty management is enhanced by services that assist users in the definition of data space access and usage controls and in the decision making required during the contract negotiation through the translation of rules in a human-understandable format. 

\subsection{Service Deployment System}
The Service Deployment system provides a marketplace designed to automate the process of preparation, packaging, and publishing of provider services in the interoperability ecosystem. Through this infrastructure, the Data Space operators and participants are able to easily browse, install, and deploy the desired modules in their existing architecture. The provided solutions leverage the containerization technologies to offer standalone solutions deployed as independent, containerized Kubernetes packages directly on the host infrastructures, aiming to reduce significantly the onboarding barriers.

\section{Agricultural Cross-Sector Data Ecosystem} \label{sec:AgroUseCase}
\begin{figure*}[!t]
\centering
\includegraphics[width=\textwidth]{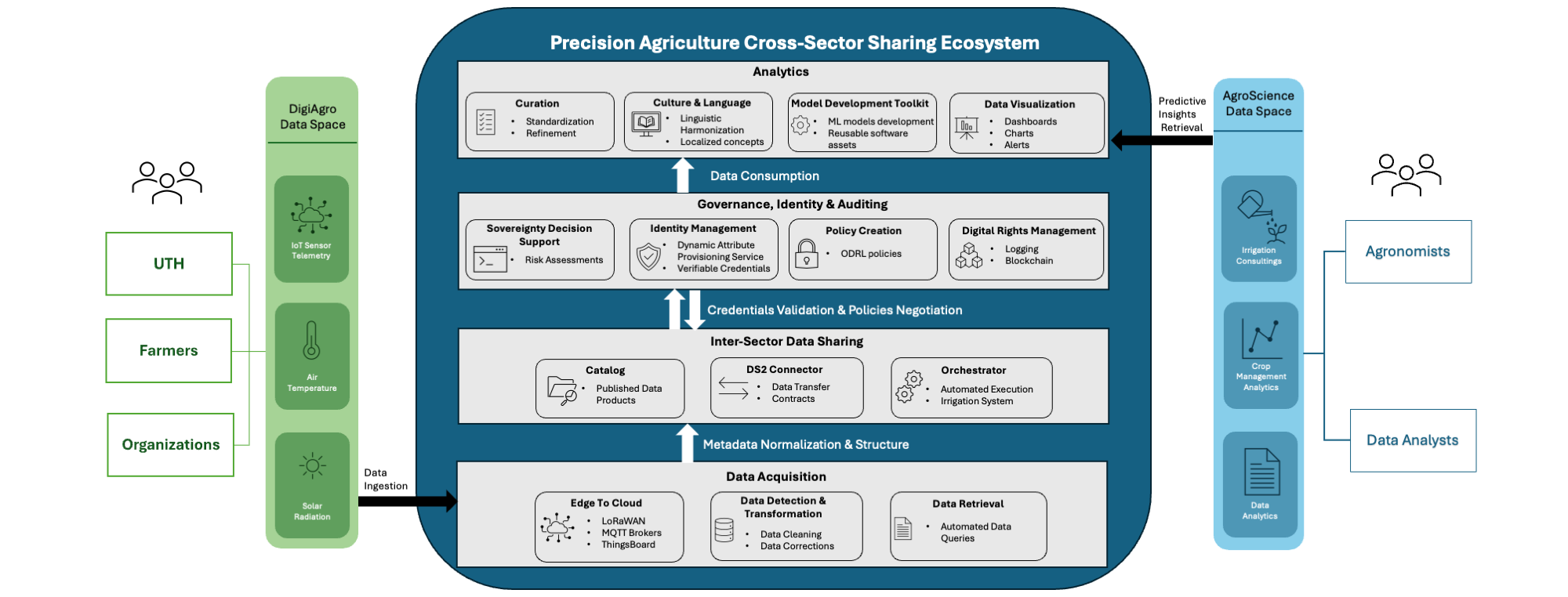}
\caption{Precision Agriculture Cross-Data-Space Ecosystem}
\label{fig:architecture}
\end{figure*}
%% Problem & Purpose 
In this section, the DSIA system is integrated in a Precision Agriculture use case to promote data-driven decision-making in existing organisations and fulfill the requirements outlined in Section \ref{sec:requirements}. The purpose of the applied framework depicted in Figure \ref{fig:architecture} is to enable agricultural, environmental, weather, and satellite data sharing among participants of different data spaces with legacy systems operating for diverse purposes, towards the improvement of crop productivity, management and sustainability. To evaluate the interoperability architecture framework, the Agricultural Cross-Sector Data Ecosystem involves two independent data spaces deployed in Northern Greece. 

The first one is DigiAgro Data Space, which operates among farms to collect, store, and exchange field measurements from IoT sensors observing soil moisture, solar radiation, localized temperature, crop management related information, and fruit images from cameras that monitor crop growth. The second one is AgroScience Data Space, which focuses on the implementation of advanced AI and ML models that provide accurate recommendations for decision-making in the optimization of crop practices for data analysts and agronomists. Such recommendations will vary from irrigation to crop management consultancies that extract valuable information about crop growth through color palette and detected size from the fruit images. Moreover, participants of AgroScience are also third-party organisations that offer satellite imagery, metrics, and weather forecasts.

Both above-mentioned precision agriculture data spaces are parts of a digital ecosystem named AgroNIT \cite{9937000}, which is a tool both for farmers and agronomists to maintain control of their offered data and services. The cooperation of the existing data spaces aims to leverage inter-data space sharing to improve productivity and sustainability while maintaining trustworthiness, privacy-preservation, and data sovereignty. 

The interoperability architecture for cross-data-space sharing provides a modular framework with independent components and services that can be obtained from a marketplace to adapt to the requirements of the involved organisations. In the Precision Agriculture use case core modules of the DSIA architecture will be utilized to formulate the cross-data-space environment, to facilitate the actual data transfers, to enforce usage policies and to derive insightful analytics of the system execution.

\subsection{Data Acquisition Layer}
The data acquisition phase is operationalized through modules installed, configured and deployed from the Marketplace directly to the host infrastructure to perform the collection and standardization of environmental data originating from farm sensor measurements. The process consists of three modules that cooperate for the purpose of the integration of raw IoT data into the data space resources. The acquisition of field telemetry is initiated by the Edge to Cloud (E2C) module,  which operates on the edge to intercept high-volume LoRaWAN streams of soil moisture, air temperature, and solar radiation from external MQTT brokers deployed on farms. E2C captures live MQTT streams ingested through the ThingsBoard IoT platform, which provides endpoints for time-series data. In DigiAgro Data Space, University of Thessaly acts as a delegated data provider and Data Space operator, after signing formal agreements with farmers allowing the publishing of their data under permitted usage policies, persisting the measurements ingested from ThingsBoard in a cloud-based TimescaleDB as time-series. 

Moreover, to access data originating from external sources, the Data Retrieval Module (RET) manages to reduce the technical complexities while enabling access to  third-party provider datasets, such as satellite imagery or weather forecasts. Its core functionality is to safely expose REST API queries from a natural language request originating from data space consumers, to retrieve external API datasets at runtime. In AgroScience Data Space, satellite metrics are managed and stored by third-party providers in PostgreSQL databases. Once data has been retrieved and normalised by E2C, it is provided to the Data Detection and Transformation (DDT) module to be analysed, cleaned, and corrected through anomaly and redundancy elimination \textbf{(FR3)}. The DDT module ensures data quality validation aiming to minimize the risk of database corruption by malformed sensor and satellite data. 

\subsection{Inter-Sector Data Sharing Layer}
The collected resources from the data acquisition phase are available for cross-data-space sharing aiming to leverage the diverse sensor data of farms to train and optimize ML models for the generation of agronomist-oriented recommendations. To achieve the inter-sector data exchange, the proposed framework utilizes the Catalog module and the DSIA Connector to enable discovery, negotiation, and peer-to-peer transactions, and automates the process by integrating the Orchestrator module. Initially, resources including IoT telemetry data or weather measurements are published to the Catalog (CAT) component as metadata listings with determined data models for offers, searches, and transactions across data spaces. Once published in the catalog, data offerings can be discovered by the verified participants of DigiAgro and AgroScience Data Spaces \textbf{(FR1)} and the data sharing process between the two data spaces relies on the DS2 Connectors deployed on both the provider and the consumer sides. To ensure secure data transfer \textbf{(TR5)} and privacy preservation of the two agricultural data spaces, automated contract agreements are initiated and mutually accepted throughout the process, while usage policies are evaluated on-the-fly \textbf{(FR2)}. 
% Moreover, automated workflows are executed by the Orchestration component which designs and handles multi-step processes across the two Data Spaces. One of the automated processes is the irrigation system, where the ORC module manages an automated cron transfer request which periodically collects IoT measurements from DigiAgro Data Space and creates contracts to feed them in the ML models of AgroScience Data Space, invoking predictive analytics regarding estimations about water requirements. 

\subsection{Governance, Identity \& Logging Layer}
While the data sharing modules efficiently perform the data bits transfer across distinct data spaces, the underlying trust verification, regulatory compliance, and security must be enhanced through purpose-specific components. Hence, additional modules are integrated in the suggested framework including federated identity, usage policy enforcement, and immutable logging management. Before the transaction execution, the Sovereignty Decision-Support (SDS) module performs risk and vulnerability assessment to establish the security baselines for the cross-data-space data sharing and produces a verifiable credential that will be utilized from the policy enforcement module to enable the data exchange process \textbf{(FR4)}. To further enhance trust in the data exchange framework, the creation and validation of identities are handled by the Identity Module (IDM) relying on existing Dynamic Attribute Provisioning Service (DAPS) \cite{omejdn_daps_2022} profiles of each Data Space and extending the identities with Verifiable Credentials \textbf{(TR2)}. The functionality of this module is linked to the DS2 Connector to enable federation in the identity management of different data spaces.

Data sovereignty is fine-tuned through the Policy Creation component (PCR), which acts as the authority that creates policy descriptions in ODRL to ensure that the data access requests comply with the rulebooks of the data spaces \textbf{(OR1, OR3)}. The module compares usage control policies of data spaces at runtime to execute and enforce them while ensuring governance \textbf{(TR3)}. Auditing is also significantly important in the cross-data-space ecosystem to protect farmer sovereignty and provide transparency on IoT data access requests and transactions. The logging in the inter-data space agricultural ecosystem is performed by the Digital Rights Management (DRM) module, which utilizes a containerized Hyperledger Fabric network to log the desired events onto an immutable blockchain ledger. The registered events include contract negotiations, orchestration workflows, and data transfers \textbf{(TR1)}. 

\subsection{Onboarding \& Deployment Layer}
In order for DigiAgro and AgroScience Data Spaces to perform inter-sector data exchange under secure circumstances, each Data Space is obliged to onboard to the DSIA framework, to be able to deploy the stack of containerized modules across heterogeneous infrastructures. For the onboarding in the ecosystem, the Portal module, acting as a cloud Software-as-a-Service (SaaS), is the necessary administrative entry point where new organisations and data spaces are verified, registered to the system and assigned to authorized roles. After onboarding to the interoperability ecosystem, the registered administrators of the participant data spaces can directly access a centralized Data Marketplace (DMK) which exposes a catalog of inter-sector published data products, such as IoT farm data and crop growth monitoring measurements, analytics and prediction services, such as recommendation ML services, and software modules to be deployed in the framework \textbf{(TR6)}. 

Upon service acquisition from the marketplace, the infrastructures of both agricultural data spaces initialize the Inter-sector Dataspace Toolkit (IDT) module which is deployed in the organisations’ local premises. This enables a host environment that executes the interoperability software modules in a scalable, data space agnostic manner through Kubernetes and containerized applications. The final deployment stage is executed after the establishment of the IDT module to minimize the installation DevOps barriers by an additional Containerisation Module (CONT) that acts as an automated tool for installation, packaging and runtime configurations of the interoperability modules.

\subsection{Analytics Layer}
The proposed framework provides a layer for insightful analytics that delivers personalized decision-support to farmers by transforming raw IoT farm data and crop growth monitoring data into user-oriented intelligence. The component responsible for the transformation and refinement of the farm datasets into standardized formats is the Curation Module (CUR). The standardized data products from the CUR module are further transformed by the Culture and Language Module (CLM) by automated linguistic harmonization in terms of vocabulary translations and localized concept mappings into a uniform structure before being forwarded to the algorithms \textbf{(TR4)}. 

The standardized data products are fed into the Model Development Toolkit module (MDT) which offers a framework where agronomists and data analysts can design, develop, and evaluate the ML models related to prediction of crop management and water requirements by packaging the algorithms as reusable software assets. Finally, the generated recommendations are forwarded to the Data Visualisation module (DVM), which offers a user interface tool including dashboards, charts, and notifications related to recommendations and crop management actions for farmers.

\section{Precision Agriculture Cross-Data-Space Data Sharing} \label{sec:workflow}

\begin{figure*}[t]
    \centering
    \includegraphics[width=\textwidth]{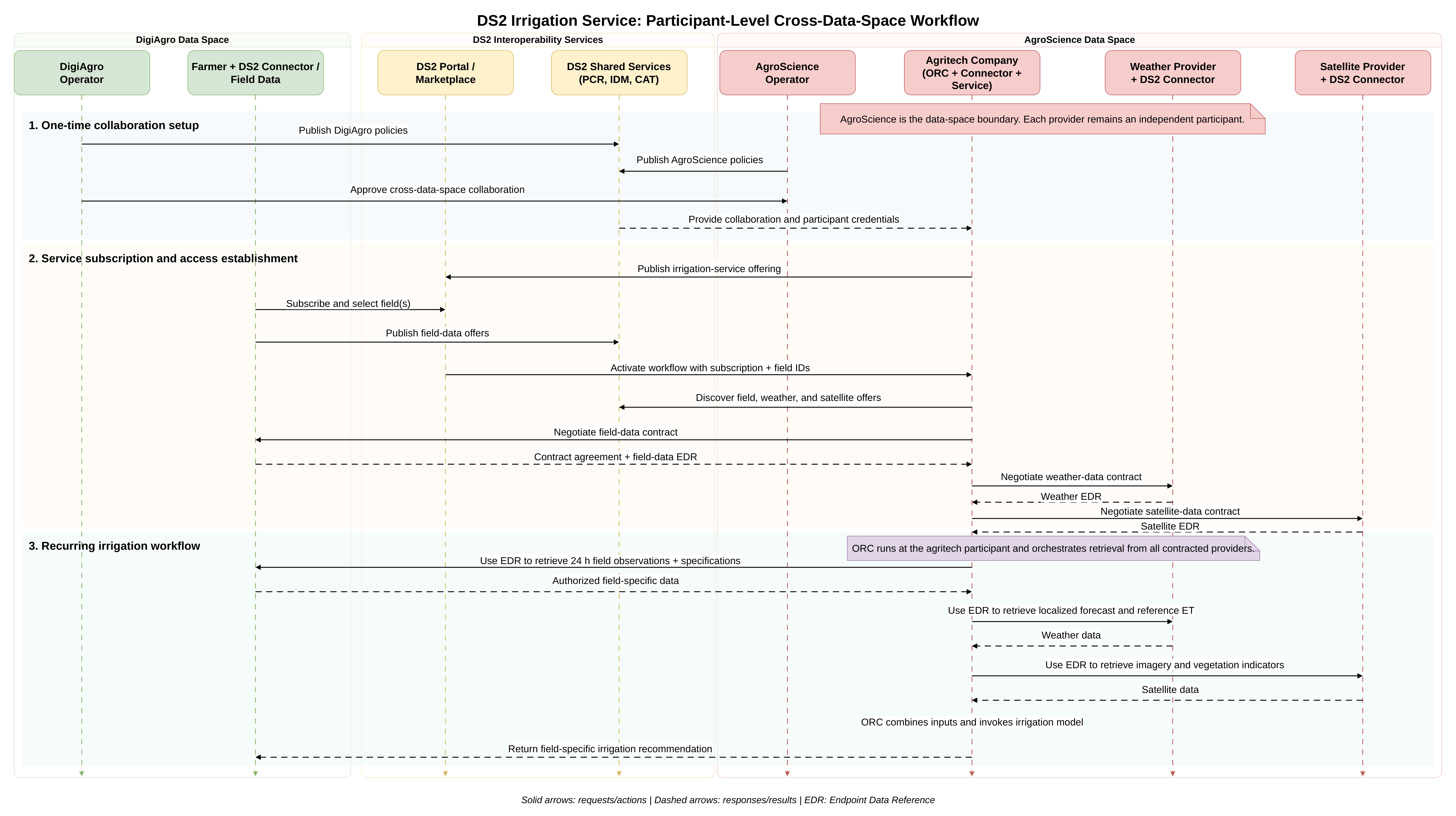}
    \caption{Precision Agriculture Cross-Data-Space Ecosystem Data Sharing Workflow}
    \label{fig:workflow}
\end{figure*}

In this section we provide the details of the implementation of a representative scenario in Precision Agriculture that leverages the interoperability framework of DS2 to perform cross-data-space exchanges between the independently governed DigiAgro and AgroScience data spaces, as shown in the sequence diagram in Figure \ref{fig:workflow}. These exchanges allow the development of a data-driven service providing field-specific irrigation recommendations.

Before the irrigation service can be executed, however, the operators of the two data spaces, along with the DSIA operator, complete a one-time cross-data-space setup. Specifically, the policies governing each data space are described through the operator-hosted Policy Creation Module (PCR) and stored in the DS2 policy repository, while the PCR-COMPARE subcomponent of the PCR implementation evaluates whether the rules of the two data spaces are compatible for the intended purpose of irrigation optimization. 

Following this evaluation and the acceptance of its outcome by the involved operators, a cross-data-space collaboration policy and a corresponding DigiAgro--AgroScience collaboration credential are created. This collaboration authorizes participants to discover data products, negotiate their terms, and acquire them across the two data spaces. The DS2 Identity Module associates the relevant participant, data-space, connector, and collaboration credentials with the authorized participants. These credentials can subsequently be retrieved from the IDM and stored in the participants' local credential wallets and DS2 Connectors.

The operational scenario is initiated by a farmer registered in the DigiAgro Data Space who provides data coming from their fields as data products. Each field is associated with a unique identifier that enables the irrigation service to access only the fields explicitly selected by the farmer. The irrigation decision-support service is provided by an agritech organisation participating in the AgroScience Data Space, and is published as an offering in the Data Marketplace Module (DMK). The farmer then accesses the marketplace through the Portal Module (PORTAL), purchases or subscribes to the service, and selects the fields for which it should be activated. The marketplace transaction establishes a service subscription. Following the purchase, the DMK creates a marketplace credential associated with the service offering and sends it to the global IDM. The farmer's local credential-management component subsequently retrieves this credential and stores it in the farmer's credential wallet. This credential represents the farmer's entitlement to use the irrigation service. However, it does not by itself authorize the AgroScience service provider to access the farmer's field data.

For each selected field, two field-specific data products are exposed through the DigiAgro Data Space. The first contains IoT environmental observations, including soil moisture, air and soil temperature, relative humidity, solar radiation, rainfall, leaf wetness, and wind information. The second contains field specifications such as crop type, planting date, field area, geographical boundary, soil texture, irrigation-system efficiency, and emitter flow. 

These assets are associated with data-offer and participant-level policies and are published in the DigiAgro provider Connector's local catalogue. They may also be advertised through the Catalogue Module (CAT) to support cross-data-space discovery. The CAT entries contain metadata, access descriptions, and machine-readable access conditions in ODRL format, while the underlying field data remain within the DigiAgro infrastructure. The corresponding policies are generated using the PCR and restrict access to the selected field, to the AgroScience irrigation service, for the purpose of irrigation optimization, and for the duration of the subscription.

The Orchestration Module (ORC) coordinates data retrieval on behalf of the AgroScience irrigation service. First, during initialization, ORC uses the farmer's subscription information to identify the field identifiers and data products required by the service. The CAT may initially be queried to identify the relevant provider and available data offers. ORC then instructs the agritech company's DS2 Connector to query the DigiAgro provider Connector's local catalogue and retrieve the current contract offers corresponding to the selected fields. For each selected data product, the agritech company's DS2 Connector initiates contract negotiation with the DS2 Connector on the DigiAgro side. 

Before the contract request is accepted, the IDM verifies the participating organisations, services, connectors, home-data-space memberships, and cross-data-space collaboration credentials. The Policy Agreement and Enforcement (PAE) module evaluates the provider and consumer policies under the cross-data-space collaboration agreement {\textbf(OR2)}. When negotiation is successfully completed, a contract agreement is established for each field-specific data product and an identifier referencing the agreement is produced. 

The consumer Connector then initiates the corresponding transfer process, which produces an Endpoint Data Reference (EDR) containing the data-plane endpoint and the authorization information required to access the contracted assets. ORC stores the EDRs of the involved data assets together with the contract-agreement identifiers and field identifiers, so that during recurring execution it does not repeat catalogue discovery and contract negotiation while the existing agreements and EDRs remain valid. Instead, it uses them to retrieve the data required by the irrigation service, which include the preceding 24 hours of IoT measurements and the corresponding field specifications. On the DigiAgro side, the Edge-to-Cloud Module (E2C) transfers measurements from the field’s LoRaWAN and MQTT infrastructure to the IoT platform, while the Data Detection and Transformation Module (DDT) validates the payloads, detects malformed values, and converts valid observations into the required representation. 

The retrieved DigiAgro data are subsequently combined with complementary resources that are obtained through AgroScience using the geographical boundary of the selected field. These resources include localized weather forecasts, forecast precipitation, reference evapotranspiration, satellite imagery, and vegetation indicators such as NDVI, EVI, and leaf-area index. The combined input is then submitted to the irrigation-estimation service. If an EDR expires, is revoked, or becomes invalid, ORC requests a renewed EDR or reinitiates the required transfer and contract procedures before continuing the workflow.

The irrigation service estimates crop water requirements by combining field measurements with meteorological and satellite-derived information. Satellite observations are used to represent the current crop condition and to estimate an evapotranspiration fraction, \(ETrF\). The daily crop evapotranspiration is then calculated as

\[
ET_{c,\mathrm{daily}} = ETrF \times ETr_{\mathrm{daily}},
\]

where \(ETr_{\mathrm{daily}}\) represents daily reference evapotranspiration derived from meteorological observations and an AgroScience reference-weather service.

Alternatively, when satellite metrics are not available, for example when there is cloudiness, crop evapotranspiration may be expressed using a crop coefficient:

\[
ET_c = K_c \times ET_0,
\]

where \(K_c\) is a crop coefficient estimated from the crop-development stage, and \(ET_0\) is the reference evapotranspiration.

The net irrigation requirement is determined by accounting for water already available to the crop. Effective rainfall is subtracted from crop evapotranspiration, while field soil-moisture measurements are used to estimate the change in water stored within the crop root zone. The net requirement may therefore be expressed as

\[
IR_n = ET_c-\left(P_e+\Delta S\right),
\]

where \(P_e\) is effective rainfall and \(\Delta S\) represents the contribution of soil-water storage. Forecast rainfall is also considered so that irrigation is reduced or postponed when sufficient precipitation is expected.

The daily net requirement may be accumulated over a soil-dependent irrigation horizon. For example, a shorter interval may be used for light or sandy soils, whereas a longer interval may be applied to medium- or heavy-textured soils with greater water-retention capacity. The accumulated requirement is subsequently adjusted according to the efficiency of the installed irrigation system and any leaching requirement associated with water salinity:

\[
I_{da}=\frac{IR_{n,\mathrm{total}}}{E_a(1-LR)},
\]

where \(I_{da}\) is the gross irrigation depth, \(E_a\) is the application efficiency, and \(LR\) is the required leaching fraction.

The corresponding water volume is calculated from the irrigation depth and the cultivated area of the selected field. The irrigation duration is then derived from the application rate of the installed system:

\[
I_{dh}=\frac{q n}{S_tS_r},
\qquad
I_t=\frac{I_{da}}{I_{dh}},
\]

where \(q\) is the emitter flow, \(n\) is the number of emitters per plant, \(S_t\) is the plant spacing, \(S_r\) is the row spacing, and \(I_t\) is the recommended irrigation duration. Information regarding the soil texture, emitters' setup and spacing of plants is extracted from the field's specifications data asset.

The AgroScience-based irrigation decision-support service returns a result associated with the specific farmer and field. The response includes crop and reference evapotranspiration, estimated crop coefficient, recommended irrigation depth, total water volume, proposed starting time, irrigation duration, and a confidence value.

The recommendation is presented to the farmer through the AgroNIT interface. The result explicitly identifies the field to which it applies and summarizes the observations and forecasts that influenced the calculation. The farmer remains in the decision loop and may approve, modify, postpone, or reject the recommendation. If the farmer cancels the marketplace subscription, removes a field from the service, or revokes the relevant policy authorization, the corresponding contract agreements are terminated, preventing subsequent workflow executions from accessing that field's data.

\section{Conclusions} \label{sec:conclusions}
This paper presents an interoperability architecture for Precision Agriculture systems with the purpose of interconnecting existing data spaces and supporting inter-data space seamless data sharing. The framework relies on bridging the DigiAgro and AgroScience Data Spaces through an interoperability ecosystem that leverages modular independent technologies that orchestrate field data collection from edge IoT devices deployed on farms and enable the generation of recommendations regarding accurate irrigation requirements. The proposed framework promotes intelligence in decision-making of crop management practices to optimize production yield and ensure sustainability. 

Future efforts will emphasize on conducting a dedicated empirical performance and scalability evaluation to quantify transaction latencies and computational resource consumption across IoT devices deployed in fields in Northern Greece. Additionally, future work will focus on enhancing semantic interoperability across heterogeneous providers by introducing standardized ontologies, such as SAREF4AGRI, while extending the proposed framework to interconnect with external European initiatives, such as the Common European Agricultural Data Space.

\bibliographystyle{IEEEtran}
\bibliography{digiagrods_revision2}

\end{document}